\documentclass[journal]{IEEEtran}
\usepackage{ifpdf}

\ifCLASSINFOpdf
   \usepackage[pdftex]{graphicx}
\else
\fi
\usepackage{amssymb,amsmath}

\newcommand{\R}{\mathbb{R}}
\DeclareMathOperator{\const}{const}

\begin{document}
%
\title{Energy Efficient Control of an Induction Machine under Load Torque Step Change}
%
%
%

\author { Alex~Borisevich,
Gernot~Schullerus

\thanks{Preliminary version}}


%
%

\markboth{Journal of \LaTeX\ Class Files,~Vol.~6, No.~1, January~2014}%
{Shell \MakeLowercase{\textit{et al.}}: Bare Demo of IEEEtran.cls for Journals}
%



\maketitle

\begin{abstract}

Optimal control of magnetizing current for minimizing induction motor power losses during load torque step change was developed. Original problem was slightly simplified and exactly optimal control was obtained for modified formulation. Obtained strategy has feedback form and is exactly optimal for ideal speed controller performance and absence of saturation in motor. Then original and simplified problem was compared in terms of quality of result and little difference found. The impact of limited bandwidth of real speed controller is analyzed. For case of main induction saturation the sub-optimal optimal control is suggested. Relative accuracy of sub-optimality is studied. Hardware implementation of optimal strategy and experimentation conducted with induction motors under vector control.

\end{abstract}

\begin{IEEEkeywords}
induction machine, power losses, dynamic operation, Pontryagin principle 
\end{IEEEkeywords}

%
\IEEEpeerreviewmaketitle

\section{Introduction}

\IEEEPARstart{T}{he} induction machine is widely used in industrial applications due to its robustness and its low cost compared to permanent magnet synchronous machines. However, in part load operation the efficiency of the induction machine dramatically decreases when the flux is kept at the nominal level. To address this issue different strategies \cite{1}-\cite{2} have been developed in the past to increase the efficiency of the induction machine in a large operation range.

The main idea of these methods is to choose an appropriate value for the rotor flux depending of the machine load in stationary state of the machine. This covers a large part of applications like pumps, fans or conveyor applications. However, when the induction machine is operated under changing loads these methods will not yield the maximum efficiency.

A practical extension of the steady-state schemes is to switch
between loss minimization control and minimum time control
depending on the actual reference \cite{3}. During a torque
transient, loss minimization is deactivated, and a minimum-time controller (similar to deadbeat control) is activated. This
attenuates the problem of slow torque response under reduced
flux magnitude. Here, also, loss minimization is not obtained
during transients.

Nevertheless, only a relatively small number of works have
addressed true loss minimization for dynamic operation. The
first treatment of this problem in \cite{4} is a purely numerical solution, and it assumes full knowledge of the speed and
torque trajectories of the application. Using offline optimization on a PC, a time-varying rotor flux trajectory is calculated,
which minimizes the controllable losses \cite{1}. The optimal trajectory is then uploaded to the controller. The experimental study
shows that this method obtains considerable loss improvements
compared to constant norm flux operation in servo applications.
However, the offline optimization is a limitation, as the optimal
flux trajectories are only valid for one specific application.

An analytic study of the problem of loss minimization during transient operation is found in \cite{5}. Here, the conditions
of optimality to minimize the total energy losses while satisfying torque-tracking constraints are calculated using calculus of
variations. However, as no analytical solution is known for this
nonlinear problem, the results for the steady state are generalized in a straightforward manner. Solving the dynamic optimization problem requires numerical algorithms. Today, however, computational power has expanded such that a numerical
scheme can be applied online.

Another solution presented in \cite{7} and consists of simple controller with cascaded structure, but the values of torque load should be known for calculations of optimal trajectory.

Recent paper \cite{6} extends \cite{5} by proposing a loss minimization
scheme that considers the dynamic problem and that is simple
enough to operate at the high sampling rates necessary in electrical drives. However, this work is not the ultimate solution. The authors relied on flux estimation and thus question of estimator arises. No mention about main inductance saturation, but it is essential for most of motors. Also only assumed that ideal speed controller is used.

The motivation of current paper is to give simple and easy implementable strategy for efficiency optimization in one particular case, the step change of motor load torque. Article material differs from previously published in following: the optimal current trajectory has a strictly feedback form, torque estimation is not needed, the trajectory is exactly optimal under natural conditions, main core saturation is taken to account as well as limited performance of speed controller has been analyzed.

From application point of view, load step change trajectory is not only could be result of mechanical load disturbances, but also occurs when reference speed trajectory is a ramp with constant acceleration and deceleration. Thus, given solution covers essential part of induction motor applications.

\section{Background}

\subsection{Motor model}

Consider the $\Gamma$-inverse model with the orientation of the rotor flux $\phi_r$ along the d-axis of synchronously rotating coordinate system \cite{8}. In the state-space model is realized by the fourth order system of differential equations:

\begin{equation}\label{eq:motor}
\begin{gathered}
\dot \phi_r = -\frac{R_R}{L_M} \phi_r  + i_{sd} R_R \\  
\dot i_{sq} = - \frac{\omega}{ L_\sigma } \phi_r - \frac{R_s}{L_\sigma} i_{sq} - \frac{R_R}{L_\sigma} i_{sq} - i_{sd} \omega_s + \frac{u_{sq}}{L_\sigma} \\
\dot i_{sd} = \frac{R_R}{L_M L_\sigma} \phi_r - \frac{R_R}{L_\sigma} i_{sd} - \frac{R_s}{L_\sigma} i_{sd} + i_{sq} \omega_s + \frac{u_{sd}}{L_\sigma}  \\
\dot \omega = p\frac{p \phi_r i_{sq} - T_m}{ J }
\end{gathered}
\end{equation}

where $\omega_s = \omega + \dfrac{R_R i_{sq}}{\phi_r}$ is a synchronous speed, $\omega$ is shaft electrical rotation speed, $T_e = p \phi_r i_{sq}$ is an electromagnetic torque produced by the motor.

Note, that the currents and voltages in model \eqref{eq:motor} are measured using power-invariant scaling of Park-Clarke transforms.

During all material of paper we will neglect the dynamics of $i_{sd}$ and $i_{sq}$ current regulators with assumption that its performance sufficiently faster than flux and speed dynamics. In this case, we can write the reduced motor model

\begin{equation}\label{eq:motor_reduced}
\begin{gathered}
\dot \phi_r = -\frac{R_R}{L_M} \phi_r  + i_{sd} R_R \\  
\dot \omega = p\frac{p \phi_r i_{sq} - T_m}{ J }
\end{gathered}
\end{equation}

which is subject of study in present work.

\subsection{Power losses model}

From model \eqref{eq:motor} it is possible to determine the power that is dissipated at $R_s$ and $R_R$, the active elements of the equivalent circuit. Directly from the physical meaning of the equivalent circuit, we obtain:

\begin{equation}\label{eq:P_dyn_pre}
P_{dyn} = R_s \cdot i_s^2 + R_R \cdot i_r^2
\end{equation}

Since the magnetic system of motor is assumed linear, then $\phi_r = L_M (i_s + i_r)$ (equation (2) in \cite{8}). On the other hand, $\phi_{rq} = 0$ when perfect orientation of the coordinate system to the rotor field is achieved. Hence we obtain the rotor currents: $i_{rd} = (\phi_r - L_M i_{sd}) / L_M$ and $i_{rq} = -i_{sq}$. Substituting these to \eqref{eq:P_dyn_pre}, we obtain

\begin{equation}
\begin{gathered}
P_{dyn} = R_s \cdot (i_{sd}^2 + i_{sq}^2) + \\
+ R_R \cdot  \left ( i_{sd}^2 + i_{sq}^2 + \frac{\phi_r^2}{L_M^2} - 2 \frac{\phi_r i_{sd}}{L_M} \right )
\end{gathered}
\end{equation}

Expression of $P_{dyn}$ can be split into two terms:

\begin{equation}
\begin{gathered}
P_{dyn}(t) = P_{loss}(t) + \Delta P(t) \\
P_{loss}(t) = i_{sq}^2(t) (R_s + R_R) + i_{sd}^2(t) R_s \\
\Delta P(t) = R_R \left ( i_{sd}(t) - \frac{\phi_r(t)}{L_M} \right )^2
\end{gathered}
\end{equation}

where $\Delta P(t)$ is an instantaneous power loss associated with the action of the current $i_{rd}$, and $\Delta P(t) \to 0$ with $t \to \infty$.

Here and after we will use $P_{loss}(t)$ for optimization criterion, skipping effect of the current $i_{rd}$. We will obtain optimal control for energy of $P_{loss}(t)$, and then demonstrate that ignoring the $\Delta P(t)$ is an acceptable compromise in accuracy.

\subsection{Optimality in steady state}

Minimum of power loss in steady state can be found by solving the equation $\partial P_{loss} / \partial i_{sd} = 0$ as a condition for a local minimum by $i_{sd}$ (\cite{1}-\cite{2}). After simplification, we obtain the optimal magnetizing current as a function of load torque $T_m$:

\begin{equation}\label{eq:i_sd_opt_ss}
i_{sd}^{opt}(T_m) = \sqrt{\frac{T_m}{L_M p}} \sqrt[4]{\frac{R_R + R_s}{R_s}} = \sqrt{\frac{T_m}{p \gamma L_M}}
\end{equation}

where $\gamma = \sqrt{\dfrac{R_s}{R_R + R_s}}$.

Since in the steady-state quadrature current is determined from output torque equation as follows $i_{sq} = \dfrac {T_m} {p L_M i_{sd}}$, we can compute the corresponding optimal value $i_{sq}^{opt}$. Trivial simplification give

\begin{equation}\label{eq:gamma}
\frac{i_{sq}^{opt}}{i_{sd}^{opt}} = \sqrt{\frac{R_s}{R_R + R_s}} \doteq \gamma
\end{equation}

Thus the ratio of the two optimal currents $i_{sq}^{opt} / i_{sd}^{opt}$ does not depend on the load torque, and depends only on the ratio of motor resistances.

\section{Optimal control problem}

The objective of the paper is optimization of motor efficiency in situation when the load changes by step from $T_m$ to $T_m + \Delta T_m$.

Suppose the speed controller is fast enough to accommodate the sudden change of torque load and the speed drop is close to zero. Then it is possible to neglect the transient processes of speed PI-controller and assume that the regulator always maintains appropriate value of qudrature current to ensure constant output torque with variation of $\phi_r$:

\begin{equation}\label{eq:speed_ctrl}
 i_{sq}(t) = \frac{T_m+\Delta T_m}{p \phi_r(t)} 
\end{equation}

and power losses with such speed controller can be expressed as

\begin{equation}\label{eq:P_loss}
P_{loss}(t) = \left ( \frac{T_m+\Delta T_m}{p \phi_r(t)} \right )^2 (R_s + R_R) + i_{sd}^2(t) R_R
\end{equation}

Thus we can setup optimal control problem with introducing cost function as an integral of power losses over transient time interval: 

\begin{equation}\label{eq:cost}
J = \int_0^T P_{loss}(t) ~ dt
\end{equation}

where $T$ is a duration of transient.

Optimal control trajectory for $i_{sd}^*$ is a minimizer for $J$ subject to flux dynamics $\phi_r$:

\begin{equation}\label{eq:flux}
\dot \phi_r = -\tau_R^{-1} \phi_r + R_R i_{sd}
\end{equation}

and boundary conditions for the state $\phi_r(0) = L_M \cdot i_{sd}^{opt}(T_m)$, $\phi_r(T) = L_M \cdot i_{sd}^{opt}(T_m+\Delta T_m)$.

\subsection{Optimal strategy}

The statement: optimal control trajectory is 

\begin{equation}\label{eq:optimal_control}
i_{sd}^*(t) = \frac{i_{sq}(t)}{\gamma}
\end{equation}

Proof.

Let's apply the Pontryagin's minimum principle to original problem.

Here we can define Hamiltonian

\begin{equation}\label{eq:Hamiltonian}
\begin{gathered}
H = \dot J + \lambda \cdot \dot \phi_r \\
H = \left ( \frac{T_m+\Delta T_m}{p \phi_r} \right )^2 (R_s + R_R) + i_{sd}^2 R_R \\
+ \lambda \cdot \left ( -\frac{R_R}{L_M} \phi_r + R_R i_{sd} \right ) 
\end{gathered}
\end{equation}

where $\lambda \in \R$ is a co-state for state variable $\phi_r$.

From the Pontryagin's minimum principle \cite{9}, the necessary optimality condition for trajectory $i_{sd}^*$ is a system of PDE:

\begin{equation}\label{eq:Pontryagin}
\begin{gathered}
 \frac{\partial H}{\partial \phi_r} = - \dot \lambda, \; \frac{\partial H}{\partial i_{sd}} = 0 
\end{gathered}
\end{equation}

subject to boundary conditions $\phi_r(0) = L_M \cdot i_{sd}^{opt}(T_m)$, $\phi_r(T) = L_M \cdot i_{sd}^{opt}(T_m+\Delta T_m)$. 

The conditions \eqref{eq:Pontryagin} is in fact sufficient \cite{9}, because

\begin{equation}
\frac{\partial^2 H}{\partial i_{sd}^2} =  2 R_s > 0
\end{equation}

Expanding \eqref{eq:Pontryagin} we got

\begin{equation}
\begin{gathered}
 \frac{\partial H}{\partial \phi_r} = - \frac{R_R}{L_M} \lambda - 2 \frac{i_{sq}^2}{\phi_r} (R_R + R_s) \\
 \frac{\partial H}{\partial i_{sd}} = R_R \lambda + 2 R_s i_{sd}
\end{gathered}
\end{equation}

From here we have costate and current dynamics

\begin{equation}\label{eq:dot_lambda}
\begin{gathered}
 \dot \lambda = \frac{R_R}{L_M} \lambda + 2 \frac{i_{sq}^2}{\phi_r} (R_R + R_s) \\
 i_{sd} = -\frac{R_R}{2 R_s} \lambda
\end{gathered}
\end{equation}

Because \eqref{eq:Hamiltonian} is not explicitly dependant from $t$ and final time $T$ is arbitrary but fixed, then as consequence from Pontryagin's minimum principle \cite{9}, the Hamiltonian must be a constant when evaluated on an extremal trajectory $i_{sd}^*$, that is: 

\begin{equation}\label{eq:Pontryagin_const}
 H(i_{sd}^*) = \const
\end{equation}

Substituting \eqref{eq:optimal_control} and $\lambda = -\dfrac{2 R_s}{R_R} i_{sd}$ from \eqref{eq:dot_lambda} to \eqref{eq:Hamiltonian} gives

\begin{equation}
\begin{gathered}
H = i_{sq}^2 (R_s + R_R) + \frac{i_{sq}^2}{\gamma^2} R_R \\
- \frac{2 R_s}{R_R} \frac{i_{sq}}{\gamma} \cdot \left ( -\frac{R_R}{L_M} \phi_r + R_R \frac{i_{sq}}{\gamma} \right ) = \frac{2 R_s \phi_r i_{sq}}{\gamma L_M}
\end{gathered}
\end{equation}

Simplifying further by using \eqref{eq:speed_ctrl} and value $\gamma$ from \eqref{eq:gamma} we got

\begin{equation}
\begin{gathered}
H = \frac{2 R_s \phi_r }{\gamma L_M} \frac{T_m+\Delta T_m}{p \phi_r} \\
= \frac{2 R_s}{p \gamma L_M}(T_m+\Delta T_m) = \const
\end{gathered}
\end{equation}

Because $H = \const$ for control \eqref{eq:optimal_control}, then due to \eqref{eq:Pontryagin_const} it is optimal trajectory.

The proof is completed.

\section{Suboptimality with respect to instantaneous power losses}

As mentioned earlier, the optimized energy of $P_{loss}(t)$ only approximately describes the power loss in the transient and the important is to compare it with the exact instantaneous power $P_{dyn}$ in accordance with \eqref{eq:P_loss}. For that we will consider the behaviour of the term $\Delta P(t)$ during transients. 

Let's define a dimensionless coefficient $k = (T_m+\Delta T_m)/T_m$, which expresses relative increase in the load torque change.

Let load torque was decreased in stepwise, which means $\Delta T_m < 0$ and $k < 1$. We can compare power loss in the steady state before the transition process and the amplitude of $\Delta P$ under control $i_{sd}^*(t)$ by \eqref{eq:optimal_control}. Let before transient value of current $i_{sd}^*$ is selected optimally by $i_{sd}^* = i_{sd}^{opt}(T_m)$ by \eqref{eq:i_sd_opt_ss}. Then the optimal power losses value for the load torque $T_m$ is:

\begin{equation}
P_{loss}^{prev} = \left ( \frac{T_m}{p L_M i_{sd}^*} \right )^2 (R_s + R_R) + {i_{sd}^*}^2 R_s
\end{equation}

After substitution of $i_{sd}^{opt}(T_m)$ from \eqref{eq:i_sd_opt_ss} we obtain:

\begin{equation}
\begin{gathered}
P_{loss}^{prev} = T_m \frac{R_s + \gamma^2 R_R + \gamma^2 R_s}{\gamma L_M p} \\
= \frac{2 R_s T_m}{p \gamma L_M}
\end{gathered}
\end{equation}

Let's find the peak value of $\Delta P(t)$. Because of $\Delta P(t)$ is everywhere decays it is sufficient to find $\Delta P(0)$, which can be expressed directly:

\begin{equation}
\begin{gathered}
\Delta P(0) = R_R \left ( \frac{T_m+\Delta T_m}{p L_M i_{sd}^* \gamma} - i_{sd}^* \right )^2 \\
= \frac{R_R T_m (k - 1)^2}{p \gamma L_M}
\end{gathered}
\end{equation}

Express the ratio of $\Delta P(0)$ to $P_{loss}^{prev}$ it is possible to write down following:

\begin{equation}
\frac{\Delta P(0)}{P_{loss}^{prev}} = \frac{R_R (k - 1)^2}{2 R_s}
\end{equation}

In the worst case when $k = 0$, which corresponds to the complete removal of mechanical load, the ratio of $\Delta P(t)$ peak to steady state value of power losses is $R_R/(2 R_s)$. Since $R_R$ and $R_s$ are usually of the same order, then we can assume that in the worst case peak of $\Delta P(t)$ is approximately two times less than the steady state power before the load torque change.

Now we turn to another variant of load torque change. Suppose torque was increased, which means that $\Delta T_m > 0$ and $k > 1$. In that case, it makes sense to compare the peak value of $P_{dyn}(t)$ with a peak value of $\Delta P(t)$, i.e., $P_{dyn}(0)$ with $ \Delta P(0)$. Direct calculations give:

\begin{equation}
\begin{gathered}
P_{dyn}(0) = P_{loss}(0) + \Delta P(0) \\
 = \frac{2 R_s T_m k^2}{p \gamma L_M} + \frac{R_R T_m (k - 1)^2}{p \gamma L_M}
\end{gathered}
\end{equation}

Taking the ratio of $\Delta P(0)$ to $P_{dyn}(0)$ after some simplifications gives following:

\begin{equation}
\frac{\Delta P(0)}{P_{dyn}(0)} = \frac{R_R (k - 1)^2}{R_R + R_R k^2 + 2 R_s k^2 - 2 R_R k}
\end{equation}

In the worst case when $k \to \infty$, which corresponds to an abrupt change of load torque from 0 to nominal. In such particular scenario the ratio of $\Delta P(t)$ peak with respect to the peak of power losses is calculated as limit:

\begin{equation}
\lim_{k \to \infty} \frac{\Delta P(0)}{P_{dyn}(0)} = \frac{R_R}{R_R + 2 R_s}
\end{equation}

For a complete picture of the difference between the two optimization objectives $P_{dyn}(t)$ and $P_{loss}(t)$ if is possible to consider two optimization problems:

1. Exact optimization problem based on $P_{dyn}(t)$ with objective $J^\circ = \int_0^T P_{dyn}(t) dt$. Solution of this problem we will denote as $i_{sd}^\circ(t)$.

2. Approximate optimization problem with objective $J^* = \int_0^T  P_{loss}(t) dt$, solution of which is $i_{sd}^*(t)$, determined by \eqref{eq:optimal_control}.

The solution of the problem 1 can only be numeric. To do this, we will define the Hamiltonian for objective functional $J^\circ$ and state constraint \eqref{eq:flux}:

\begin{equation}
\begin{gathered}
H = i_{sq}^2(t) (R_s + R_R) + i_{sd}^2(t) R_s + \\
+ R_R \left ( i_{sd}(t) - \frac{\phi_r(t)}{L_M} \right )^2 + \lambda (-\frac{R_R \phi_r(t)}{L_M} + i_{sd} R_R )
\end{gathered}
\end{equation}

From the conditions of Pontryagin's minimum principle $\frac{\partial H}{\partial i_{sd}} = 0$ and $-\frac{\partial H}{\partial i_{sd}} = \dot \lambda$ (equation \eqref{eq:Pontryagin}) we can obtain boundary value problem:

\begin{equation}\label{eq:BVP_dyn}
\begin{gathered}
i_{sd} = -\frac{R_R \lambda - 2 R_R \phi_r/L_M}{2 R_R + 2 R_s} \\
\dot \lambda = \frac{R_R \lambda}{L_M} + 2 \frac{i_{sq}^2 (R_R + R_s)}{\phi_r} + 2 \frac{R_R (i_{sd} - \phi_r/L_M}{L_M}
\end{gathered}
\end{equation}

with boundary conditions $\phi_r(0) = L_M i_{sd}^{opt}(T_m)$, $\phi_r(T) = L_M i_{sd}^{opt}(T_m + \Delta T_m)$ and state dynamics \eqref{eq:flux} for $\phi_r$ and trajectory $i_{sq}$ given by \eqref{eq:speed_ctrl}.

Figures 1 and 2 show the computed trajectories of $P_{dyn}(t)$ for model of DRS112M4 motor (4 kW nominal power) under two variants of $i_{sd}(t)$ control: exact solution $i_{sd}^\circ(t)$ of boundary value problem \eqref{eq:BVP_dyn} and approximate (suboptimal) control trajectory $i_{sd}^*(t)$ given by \eqref{eq:optimal_control}. Two variants of a load torque step change: from 10 \% to 100 \% (full load) and in reverse direction from 100\% to 10\% was considered. Duration of the transient $T$ was determined as the time during which $\dot P_{dyn}(t)$ becomes less than $\epsilon \cdot \max_t | \dot P_{dyn}(t) |$, where $\epsilon = 0.1$ \%.

\begin{center}
\ifpdf 
  \includegraphics[width=0.5\textwidth]{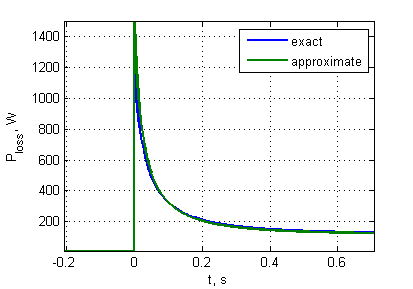}
\fi

Figure 1.a. Trajectories of $P_{dyn}(t)$ during load step change from 10 \% to 100 \% of nominal value for exact $i_{sd}^\circ(t)$ and approximate $i_{sd}^*(t)$ control.

\ifpdf 
  \includegraphics[width=0.5\textwidth]{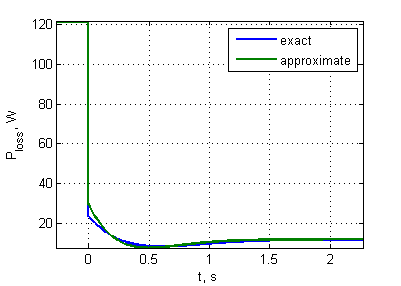}
\fi

Figure 1.b. Trajectories of $P_{dyn}(t)$ during load step change from 100 \% to 10 \% of nominal value for exact $i_{sd}^\circ(t)$ and approximate $i_{sd}^*(t)$ control.
\end{center}

For the evaluation and comparison of exact solution $i_{sd}^\circ(t)$ and approximate solution $i_{sd}^*(t)$ let's calculate two quantities: energy (integral) of $P_{dyn}(t)$ with a control $i_{sd}^\circ(t)$, i.e. $J_1 = P_{dyn}(t) \bigg|_{i_{sd} = i_{sd}^\circ(t)}$ and energy of $P_{dyn}(t)$ with a control $i_{sd}^*(t)$, i.e. $J_2 = P_{dyn}(t) \bigg|_{i_{sd} = i_{sd}^*(t)}$. In tables 1 and 2 values of energy $J_1$ and $J_2$ for motors of 0.4 kW rated power (DRS71S4), 4 kW rated power (DRS112M4) and 11 kW rated power (DRS160M4) are given. From the energies $J_1$ and $J_2$ absolute error $\Delta J = J_2 - J_1$ and relative error $\delta J = \Delta J/J_1$ for approximate solution were determined.

\begin{center}
Table 1.
Energy losses for the exact and approximate solutions to a step decrease in torque from 100 \% to 10 \% of nominal load.

\begin{tabular}{ | c | c | c | c | c | c | }
\hline 
Motor & T, s & $J_1$, J & $J_2$, J & $\Delta J$, J & $\Delta J/J_1$, \% \\
\hline 
DRS71S4  & 0.62 & 3.18 & 3.22 & 0.045 & 1.43 \\
DRS71S4  & 2.27 & 25.8 & 26.4 & 0.63 & 2.45 \\
DRS71S4  & 3.86 & 62.0 & 64.0 & 1.99 & 3.20 \\
\hline 
\end{tabular}

\end{center}

\begin{center}
Table 2.
Energy losses for the exact and approximate solutions to a step increase in torque from 10 \% to 100 \% of nominal load.

\begin{tabular}{ | c | c | c | c | c | c | }
\hline 
Motor & T, s & $J_1$, J & $J_2$, J & $\Delta J$, J & $\Delta J/J_1$, \% \\
\hline 
DRS71S4  & 0.20 & 20.52 & 20.73 & 0.21 & 1.02 \\
DRS112M4 & 0.70 & 159.8 & 162.4 & 2.62 & 1.64 \\
DRS160M4 & 1.31 & 396.1 & 403.4 & 7.22 & 1.82 \\
\hline 
\end{tabular}

\end{center}

The results of calculations show that the use of the criterion $J_2 = \int_0^T P_{loss}(t) dt$ instead of $J_1 = \int_0^T P_{dyn}(t) dt$ to determine the optimal control trajectory $i_{sd}$ gives an acceptable error (less 4 \% for considered drives), which grows only slightly with an increase in the nominal motor power.

\section{Speed controller}

Due to limited bandwidth, the behaviour of speed controller differs from ideal described by \eqref{eq:speed_ctrl}. Thus the trajectory \eqref{eq:optimal_control} is sub-optimal for system with non-ideal speed controller.

Here and we will use variant of speed controller, in which the output of the PI-controller is a torque setpoint \cite{11}. Briefly it can be described as follows.

Consider following regulator with flux-dependant nonlinear scaling

\begin{equation}\label{eq:speed_PI}
i_{sq}^{ref} = \frac{1}{p \phi_r} PI[\omega_r^{ref} - \omega_r]
\end{equation}

where $\omega_r^{ref}$ is mechanical speed setpoint, $i_{sq}^{ref}$ is a setpoint for quadrature current controller, and $PI[...]$ is an output of PI-controller for given input. 

Neglecting current controller dynamics and supposing that $i_{sq} = i_{sq}^{ref}$, the closed-loop speed dynamics transformed into form

\begin{equation}
\dot \omega_r = \frac{PI[\omega_r^{ref} - \omega_r] - T_m}{J}
\end{equation}

which is obtained after substitution of $i_{sq}$ from \eqref{eq:speed_PI} to mechanical part of motor model \eqref{eq:motor_reduced}.

Controller \eqref{eq:speed_PI} has following advantages:

- the speed dynamics is completely linear,

- the speed and torque is decoupled from flux $\phi_r$ and invariant under any trajectory of magnetizing current $i_{sd}$.

Transfer function of speed controller from $T_m$ to $T_e$ is

\begin{equation}
\frac{T_e(s)}{T_m(s)} = \frac{K_p s + K_i}{J s^2 + K_p s + K_i}
\end{equation}

Consider response of $T_e$ for load step change from $T_m$ to $T_m + \Delta T_m$. As a well-known fact from the second-order systems, the output of PI-controller and $T_e(t)$ has following analytical response

\begin{equation}\label{eq:torque_delta}
\begin{gathered}
T_e(t) = T_m + \Delta T_m + \delta(t) \\
\delta(t) = \frac{\Delta T_m}{2 \sqrt{z^2 - 1} w_0}\left ( \lambda_1 e^{\lambda_1 t} - \lambda_2 e^{\lambda_2 t} \right ) \\
\lambda_1 = (-z - \sqrt{z^2 - 1}) w_0, \; \lambda_2 = (-z + \sqrt{z^2 - 1}) w_0 \\
\end{gathered}
\end{equation}

where $w_0$ is a natural frequency and $z > 1$ is a damping factor, which are expressed in terms of coefficients of PI-controller as follows: 
$K_i = J w_0^2$, $K_p = 2 z \sqrt{J K_i}$

Comparing \eqref{eq:torque_delta} and \eqref{eq:speed_ctrl}, dynamical response of non-ideal speed controller can be written

\begin{equation}\label{eq:speed_real}
 i_{sq}(t) = \frac{T_m+\Delta T_m + \delta(t)}{p \phi_r(t)} 
\end{equation}

Speed response for load step change has following analytical form

\begin{equation}\label{eq:speed_delta}
\begin{gathered}
\Delta \omega(t) = \omega_r(t) - \omega^{ref} \\
= -\frac{\Delta T_m}{2 J \sqrt{z^2 - 1} w_0}\left ( e^{\lambda_1 t} - e^{\lambda_2 t} \right )
\end{gathered}
\end{equation}

in which amplitude of error term $\Delta \omega(t)$ is clearly depends from $\sqrt{z^2 - 1} w_0$.

\subsection{Optimality study}

Let's consider impact of non-ideality of speed controller for transient energy under control \eqref{eq:optimal_control}. 

It is pretty obvious, that \eqref{eq:speed_ctrl} is a asymptotical special case of \eqref{eq:speed_real} where $w_0 \to \infty$ and $\delta(t) \to 0$. Thus, it is meaningful to characterize dependence of transient energy from $w_0$. Let's express power losses $P_{loss}$ using equation for non-ideal speed controller \eqref{eq:speed_real}:

\begin{equation}\label{eq:P_loss_real}
\begin{gathered}
\widehat{P_{loss}}(t) = \left ( \frac{T_m + \Delta T_m + \delta(t)}{p \phi_r(t)} \right )^2 (R_s + R_R) \\
+ i_{sd}^2(t) R_R = \left ( \frac{T_m + \Delta T_m}{p \phi_r(t)} \right )^2 (R_s + R_R) \\
+ \delta(t) \frac{2(T_m + \Delta T_m) + \delta(t)}{p^2 \phi_r^2(t)} (R_s + R_R) + i_{sd}^2(t) R_R \\
= P_{loss}(t) + \delta(t) \frac{2(T_m + \Delta T_m) + \delta(t)}{p^2 \phi_r^2(t)} (R_s + R_R)
\end{gathered}
\end{equation}

The total transient energy is 
\begin{equation}\label{eq:cost_real}
\begin{gathered}
\widehat{J} = \int_0^T \widehat{P_{loss}}(t) ~ dt \\ 
= J + \int_0^T \delta(t) \frac{2(T_m + \Delta T_m) + \delta(t)}{p^2 \phi_r^2(t)} (R_s + R_R) ~ dt
\end{gathered}
\end{equation}

where $J$ is a cost function \eqref{eq:cost} for ideal speed controller, which is minimized by control rule \eqref{eq:speed_ctrl}.

The value of last integral in \eqref{eq:cost_real} depends on time interval where $\delta(t) > 0$, which is determined by $w_0$. Thus if there exist two particular values of $w_0^1$ and $w_0^2$ where $w_0^1 > w_0^2$ then $\widehat{J}^1 < \widehat{J}^2$, i.e. it is possible to give numerically the upper bound of $\widehat{J}$ considering speed controller with worst practically accepted performance and lowest $w_0$.

\begin{center}
\ifpdf 
  \includegraphics[width=0.5\textwidth]{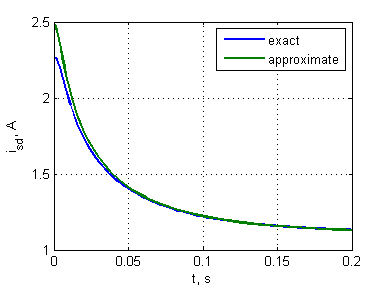}
\fi

\ifpdf 
  \includegraphics[width=0.5\textwidth]{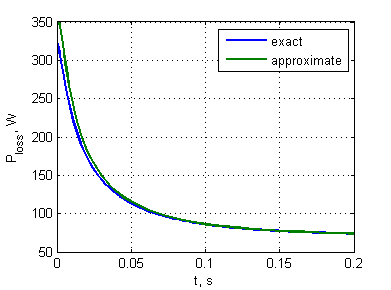}
\fi

Figure 2. Trajectories of $i_{sd}(t)$ and $P_{loss}(t)$ for optimal solutions in case of $w_0 = 20$
\end{center}

Numerical study was conducted with model of motor DRS71S4. Parameters of speed controller selected in such way that maximal speed drop was approximately 10 \% in case of step load increase from 25 \% to 100 \% of nominal torque. It gives $w_0 = 20$ and $z = 10$. Then two solutions were compared: direct numerical integration of boundary value problem \eqref{eq:dot_lambda}, \eqref{eq:flux} as a exact solution and \eqref{eq:optimal_control} as approximate one. Energy $J_1$ was calculated as total loss power during transient for exact solution. Energy $J_2$ was calculated for approximate solution. Two solutions was compared in terms of absolute $\Delta J = J_2 - J_1$ and relative $\Delta J/J_1$ errors. The duration of transient is 0.4 sec. 

The results are presented in table 3. Example of $i_{sd}(t)$ and $P_{loss}(t)$ for both solutions in case of $w_0 = 20$ presented at Figure 2.

\begin{center}
Table 3.
Transient energy for load step change from 25 \% to 100 \% of rated torque.

\begin{tabular}{ | c | c | c | c | c | }
\hline 
  $w_0$ & $J_1$, J & $J_2$, J & $\Delta J$, J & $\Delta J/J_1$, \% \\
\hline 
20 &  35.87 &  36.49  &  0.623  &  1.737 \\
40 &  35.86 &  36.17  &  0.310  &  0.865 \\
60 &  35.88 &  36.07  &  0.195  &  0.543 \\
\hline 
\end{tabular}

\end{center}

From the obtained results it is evident that for practically used speed controller parameters the relative error of approximate solutions is about 2 \% and decreases with enhancement of controller performance.

\section{Main inductance saturation}

Motor inductances are usually subject to saturation. We will model saturation as current-dependant main inductance which affects the flux dynamics

\begin{equation}\label{eq:flux_sat}
\dot \phi_r = -\frac{R_R}{L_m(i_{sd})} \phi_r + R_R i_{sd}
\end{equation}

where $L_m(i_{sd})$ is saturation curve. 

It can be shown, that model \eqref{eq:flux_sat} is compatible to general model of main inductance with nonideal core \cite{10} which magnetic flux characteristic is $\phi = f(i) = i \cdot L(i)$.

It is possible to setup optimal control problem for trajectory $i_{sd}(t)$ in case with flux dynamics \eqref{eq:flux_sat} under saturation.
New Hamiltonian has the same form as \eqref{eq:Hamiltonian}:

\begin{equation}\label{eq:Hamiltonian_sat}
\begin{gathered}
H = \left ( \frac{T_m+\Delta T_m}{p \phi_r} \right )^2 (R_s + R_R) + i_{sd}^2 R_R + \\
 \lambda \cdot \left ( -\frac{R_R}{L_m(i_{sd})} \phi_r + R_R i_{sd} \right )
\end{gathered} 
\end{equation}

According to Pontryagin's minimum principle \eqref{eq:Pontryagin}, the optimal trajectory $i_{sd}^*$ is a solution of boundary-value problem:

\begin{equation}\label{eq:BVP_sat}
\begin{gathered}
\dot \phi_r = -R_R \frac{\phi_r}{L_m(i_{sd})} + i_{sd} R_R \\
\dot \lambda = R_R \frac{ \lambda}{L_m(i_{sd})} + 2 (R_R + R_s) \frac{i_{sq}^2}{\phi_r} \\
i_{sq} = \frac{T_m + \Delta T_m}{p \phi_r} \\
R_R \lambda \left( 1 + \frac{\phi_r \frac{\partial L_m}{\partial i_{sd}}}{L_m(i_{sd})^2} \right ) + 2 R_s i_{sd} = 0
\end{gathered}
\end{equation}

subject to boundary conditions $\phi_r(0) = L_m(i_{sd}^{opt}(T_m)) \cdot i_{sd}^{opt}(T_m)$, $\phi_r(T) = L_m(i_{sd}^{opt}(T_m+\Delta T_m)) \cdot i_{sd}^{opt}(T_m+\Delta T_m)$. 

Due to nonlinear nature of \eqref{eq:BVP_sat} it is impossible to integrate BVP problem forward and produce feedback rule for optimal control. Thus approximate solutions are desirable.

\subsection{Approximate solution}

Let's consider steady-state power losses which is equals to  

\begin{equation}\label{eq:P_loss_sat}
\begin{gathered}
 P_{loss}(i_{sd}, i_{sq}) = i_{sq}^2 (R_s + R_R) + i_{sd}^2 R_s \\
 = \left ( \frac{T_m}{L_m(i_{sd}) i_{sd} p} \right )^2 (R_s + R_R) + i_{sd}^2 R_s 
\end{gathered}
\end{equation}

Differentiation gives

\begin{equation}\label{eq:dP_loss_sat}
\begin{gathered}
\frac{\partial P_{loss}}{\partial i_{sd}} = 2 R_s i_{sd} - 2 \frac{T_m^2 }{L_m^2(i_{sd}) i_{sd}^3 p^2} (R_s + R_R) \\
- 2 \frac{T_m^2 }{L_m^3(i_{sd}) i_{sd}^2 p^2} \frac{\partial L_m}{\partial i_{sd}} (R_s + R_R) \\
= 2 R_s i_{sd} - 2 (R_s + R_R) i_{sq}^2 \left ( \frac{1}{i_{sd}} + \frac{1}{L_m(i_{sd})} \frac{\partial L_m}{\partial i_{sd}} \right )
\end{gathered}
\end{equation}

Necessary and sufficient condition for minimum of steady-state power losses is 

\begin{equation}\label{eq:sat_cond}
\frac{\partial P_{loss}}{\partial i_{sd}}(i_{sd}, i_{sq}) = 0
\end{equation}

Due to last term in \eqref{eq:dP_loss_sat} it is in general impossible to solve equation \eqref{eq:sat_cond} analytically, but it can be traced numerically and we got function

\begin{equation}
i_{sd} = \zeta(i_{sq})
\end{equation}

such that condition \eqref{eq:sat_cond} is always satisfied

\begin{equation}\label{eq:zeta_cond}
\frac{\partial P_{loss}}{\partial i_{sd}}(\zeta(i_{sq}), i_{sq}) = 0
\end{equation}

Because $\eqref{eq:zeta_cond}$ is satisfied for any $i_{sq}$ as well for steady states, then $i_{sd}$ converges to optimum

\begin{equation}\label{eq:zeta}
i_{sd}(t) = \zeta(i_{sq}(t)) \to i_{sd}^{opt} = \zeta(i_{sq}^{opt})
\end{equation}

\subsection{Optimality}

Direct numerical calculations gives that the rule \eqref{eq:zeta} is not global mimimizer for \eqref{eq:cost} in case of main inductance saturation, but close to it. 

Let's denote $i_{sd}^*$ -- is exact minimizer of \eqref{eq:cost} as a solution for \eqref{eq:BVP_sat}, $i_{sd}^\circ$ -- is sub-optimal trajectory provided by \eqref{eq:zeta}: $i_{sd}^\circ(t) = \zeta(i_{sq}(t))$.

It is easy to see that $i_{sd}^\circ$ is a minimizer for system where the flux dynamics \eqref{eq:flux_sat} replaced to Wiener-type model

\begin{equation}\label{eq:flux_approx}
\begin{gathered}
\dot {\widetilde{i_{sd}}} = -\frac{R_R}{L_M} \widetilde{i_{sd}} + R_R i_{sd} \\
\widetilde{\phi_r} = L_m(\widetilde{i_{sd}}) \cdot \widetilde{i_{sd}}
\end{gathered}
\end{equation}

and the objective function replaced to 

\begin{equation}\label{eq:J_sat}
\begin{gathered}
J = \int_0^T \widetilde{P_{loss}}(t) dt \\
= \int_0^T \left ( \frac{T_m+\Delta T_m}{p \widetilde{\phi_r}} \right )^2 (R_s + R_R) + \widetilde{i_{sd}}^2 R_s dt
\end{gathered}
\end{equation}

The difference between original optimization problem for \eqref{eq:flux_sat} and modified one \eqref{eq:flux_approx}, \eqref{eq:J_sat} can be summarized as follows:

- flux dynamics divided to first-order linear dynamical element and nonlinear static map;

- energy measure $J$ depends only on $\widetilde{i_{sd}}$, thus manipulated input $i_{sd}$ is taken into account for measure through the first-order linear system.

It is evident that with reducing time rotor constant $\tau_R = L_M/R_R$ the model \eqref{eq:flux_approx} becomes closer to \eqref{eq:flux_sat}, because the flux value more precisely estimated by steady-state value $\phi_r \approx i_{sd} L_m(i_{sd})$ and dynamics of flux has less influence to overall transient energy. 

Let's introduce the suboptimality measure as a difference between total power (energy) $J$ obtained for exact optimal solution $i_{sd}^*$ for model \eqref{eq:flux_sat} and approximate $i_{sd}^\circ$ for \eqref{eq:flux_approx}. 

\begin{equation}\label{eq:delta_J}
\Delta J = \int_0^T \widetilde{P_{loss}}(t) - P_{loss}(t) dt
\end{equation}

Because in steady-state the power $\widetilde{P_{loss}}(t)$ is equal to $P_{loss}(t)$ for original model \eqref{eq:flux_sat} then the energy error is proportional to transient time, which is determined by rotor time constant $\Delta J \sim \tau_R$.

Thus it is possible to give upper bound of $\Delta J$ numerically by considering the motor with a highest rated power. We performed numerical calculation of optimal current trajectory for three different motors models with a rated power 0.4 kW (DRS71S4), 4 kW (DRS112M4) and 11 kW (DRS160M4). 
For the all motors, saturation curve is taken as a linear function: $L_m(i_{sd}) = 2 L_M + \frac{L_M}{2 i_{sd}^{nom}}$ (where $i_{sd}^{nom}$ is a nominal direct current).

First test was performed for load step increase from 25 \% to 100 \% of rated torque. The second test is step decrease from 100 \% to 25 \% of rated torque. Obtained results are summarized in table 4 and 5. Energy $J_1$ was calculated as total loss power during transient for exact solution $i_{sd}^*$ of \eqref{eq:BVP_sat}. Energy $J_2$ was calculated for approximate solution $i_{sd}^*$ by \eqref{eq:zeta}. Two solutions was compared in terms of absolute $\Delta J = J_2 - J_1$ and relative $\Delta J/J_1$ errors. Example of trajectories $i_{sd}(t)$ and $P_{loss}(t)$ for both solutions in case of load increase (first row in table 4) for motor DRS71S4 are presented at Figure 3.

\begin{center}
Table 4.
Calculated transient energy for load step change from 25 \% to 100 \% of rated torque.

\begin{tabular}{ | l | c | c | c | c | c | }
\hline 
  Motor & $\tau_R$ & $J_1$ & $J_2$ & $\Delta J$ & $\Delta J/J_1$, \% \\
\hline 
DRS71S4 & 0.065 &  51.23 &  51.45  &  0.219  &  0.427 \\
DRS112M4 & 0.238 & 365.0 & 366.2  &  1.238  &  0.339 \\ 
DRS160M4 & 0.404 & 683.2 & 685.3  &  2.149  &  0.315 \\
\hline 
\end{tabular}

\end{center}

\begin{center}
Table 5.
Calculated transient energy for load step change from 100 \% to 25 \% of rated torque.

\begin{tabular}{ | l | c | c | c | c | c | }
\hline 
  Motor & $\tau_R$ & $J_1$ & $J_2$ & $\Delta J$ & $\Delta J/J_1$, \% \\
\hline 
DRS71S4 & 0.065  & 5.62   & 5.625   & 0.005   & 0.085 \\
DRS112M4 & 0.238  & 42.1  & 42.13  &  0.034  &  0.08 \\
DRS160M4 & 0.404  & 99.32  & 99.39  &  0.07  &  0.07 \\
\hline 
\end{tabular}

\end{center}

\begin{center}
\ifpdf 
  \includegraphics[width=0.5\textwidth]{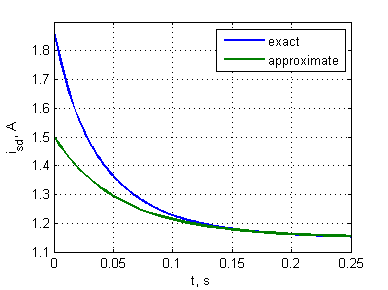}
\fi

\ifpdf 
  \includegraphics[width=0.5\textwidth]{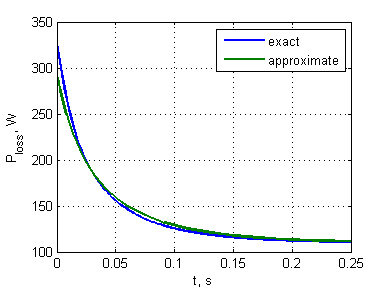}
\fi

Figure 3. Trajectories of $i_{sd}(t)$ and $P_{loss}(t)$ for exact and approximate solutions for DRS71S4 motor.
\end{center}

Conclusions from calculated results:

- with increasing of the rated motor power, the absolute error $\Delta J$ between exact and approximate solutions increases as well,

- in the same time, with increasing of the rated motor power, the relative error $\Delta J/J_1$ decreases, because transient energy $J_1$ becomes higher for bigger motors.

\section{Experimental validation}

The experiment was performed with SEW-Eurodrive DRS71S4 motor under FOC control for validate theoretically developed method. Before actual experiments, saturation curve $L_m(i_{sd})$ was identified and practically proved that $i_{sd}^{opt} = \zeta(i_{sq}^{opt})$ with acceptable accuracy (less than 4 W difference between predicted and real minimum of $P_{loss}$, Figure 4).

\begin{center}
\ifpdf 
  \includegraphics[width=0.5\textwidth]{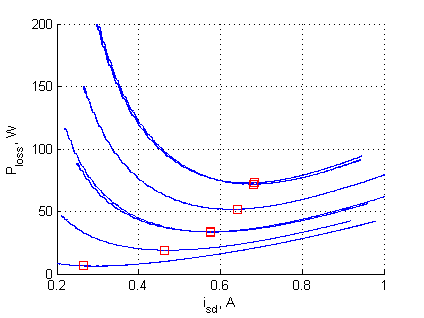}
\fi

Figure 4. Measured $P_{loss}$ for load torque from 0 to 1.42 Nm (squares are minimum points where \eqref{eq:dP_loss_sat} is 0)
\end{center}

We compared three different approaches for control $i_{sd}$ during load transient:

- "nominal": optimization is switched off, magnetizing current $i_{sd}$ is always set to nominal level $i_{sd}(t) := i_{sd}^{nom}$,

- "optimal": $i_{sd}(t) = \zeta(i_{sq}(t))$, which is subject of paper,

- "step": $i_{sd}(t) = i_{sd}^{opt}(T_m)$, which is just step change to optimal steady state value of $i_{sd}$ for given $T_m$ at beginning of transient (when load change detected), not applicable in practice because information about future value of torque $T_m$ is needed.

The experiment was conducted under step increase of load torque from 0.355 Nm to 1.42 Nm (from 14 \% to 55 \% of rated torque) and step decrease in opposite direction from 1.42 Nm to 0.355 Nm. Two speed setpoints tested: 60 rad/s and 100 rad/s (573 rpm and 955 rpm).

The transients of $P_{loss}$ are presented at Figure 5-8 for different situations. For selected transient duration $T = 0.3$ sec energy $J$ was calculated for each of tested approaches by \eqref{eq:cost}. The results are summarized in Table 6 and 7.

\begin{center}
Table 6.
Measured transient energy for load step change from 0.355 Nm to 1.42 Nm at speed 100 rad/sec.

\begin{tabular}{ | l | c | c | c | c | c | }
\hline 
Method & $J_{rise}$, 100 rad/s & $J_{fall}$, 100 rad/s \\
\hline 
nominal & 25.86  & 9.07   \\
optimal & 23.72  & 7.52   \\
step & 23.98  & 7.16   \\
\hline 
\end{tabular}

\end{center}

\begin{center}
Table 7.
Measured transient energy for load step change from 0.355 Nm to 1.42 Nm at speed 60 rad/sec.

\begin{tabular}{ | l | c | c | c | c | c | }
\hline 
Method & $J_{rise}$, 60 rad/s & $J_{fall}$, 60 rad/s \\
\hline 
nominal & 21.88  & 9.97   \\
optimal & 21.05  & 8.82   \\
step & 21.06  & 8.24   \\
\hline 
\end{tabular}

\end{center}

\begin{center}
\ifpdf 
  \includegraphics[width=0.5\textwidth]{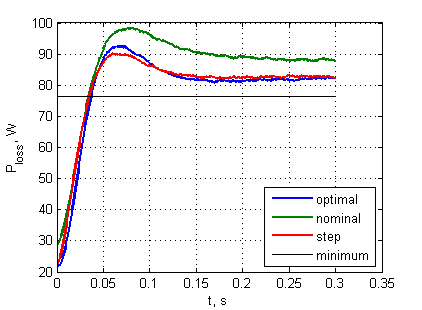}
\fi

Figure 5. Measured $P_{loss}$ for load torque step increase from 0.355 Nm to 1.42 Nm at 100 rad/s for 0.37 kW motor.
\end{center}

\begin{center}
\ifpdf 
  \includegraphics[width=0.5\textwidth]{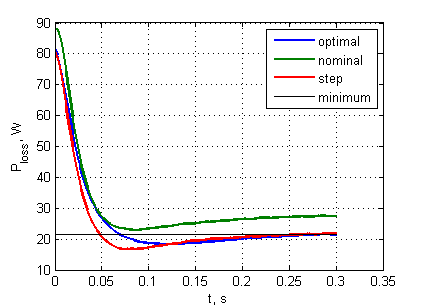}
\fi

Figure 6. Measured $P_{loss}$ for load torque step decrease from 1.42 Nm to 0.355 Nm at 100 rad/s for 0.37 kW motor.
\end{center}

\begin{center}
\ifpdf 
  \includegraphics[width=0.5\textwidth]{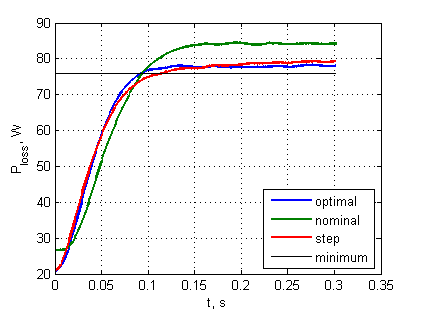}
\fi

Figure 7. Measured $P_{loss}$ for load torque step increase from 0.355 Nm to 1.42 Nm at 60 rad/s for 0.37 kW motor.
\end{center}

\begin{center}
\ifpdf 
  \includegraphics[width=0.5\textwidth]{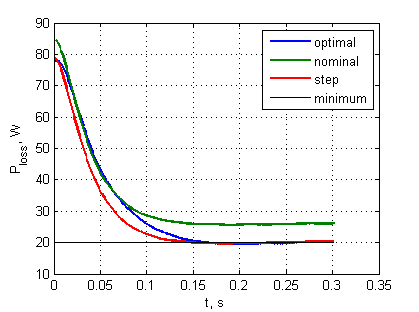}
\fi

Figure 8. Measured $P_{loss}$ for load torque step decrease from 1.42 Nm to 0.355 Nm at 60 rad/s for 0.37 kW motor.
\end{center}

The same experiment was conducted with SEW-Eurodrive DRS112M4 motor of 4 kW rated power. Step increase of load torque from 6.8 Nm to 13.6 Nm (approximately from 25 \% to 50 \% of rated torque) and step decrease in opposite direction from 13.6 Nm to 6.8 Nm are tested. Two speed setpoints tested: 60 rad/s and 100 rad/s (573 rpm and 955 rpm).

The transients of $P_{loss}$ are presented at Figure 9-12 for different situations. For selected transient duration $T = 1$ sec, energy $J$ \eqref{eq:cost} was calculated for each of tested approaches. The results are summarized in Table 8 and 9.

\begin{center}
\ifpdf 
  \includegraphics[width=0.5\textwidth]{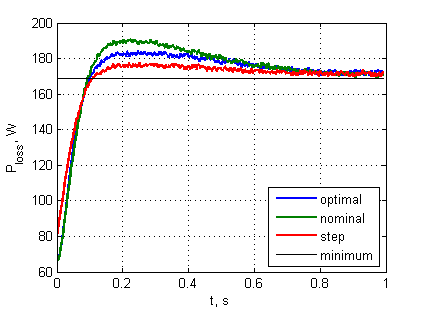}
\fi

Figure 9. Measured $P_{loss}$ for load torque step increase from 6.8 Nm to 13.6 Nm at 100 rad/s for 4 kW motor.
\end{center}

\begin{center}
\ifpdf 
  \includegraphics[width=0.5\textwidth]{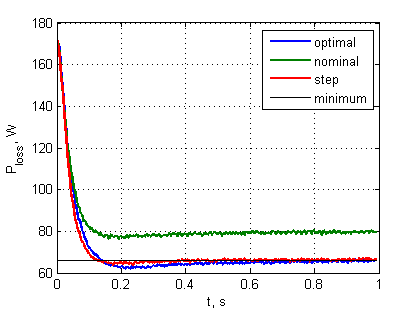}
\fi

Figure 10. Measured $P_{loss}$ for load torque step decrease from 6.8 Nm to 13.6 Nm at 100 rad/s for 4 kW motor.
\end{center}

\begin{center}
\ifpdf 
  \includegraphics[width=0.5\textwidth]{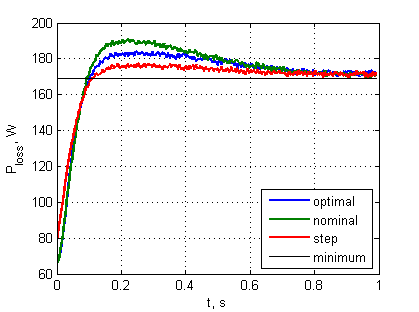}
\fi

Figure 11. Measured $P_{loss}$ for load torque step increase from 6.8 Nm to 13.6 Nm at 60 rad/s for 4 kW motor.
\end{center}

\begin{center}
\ifpdf 
  \includegraphics[width=0.5\textwidth]{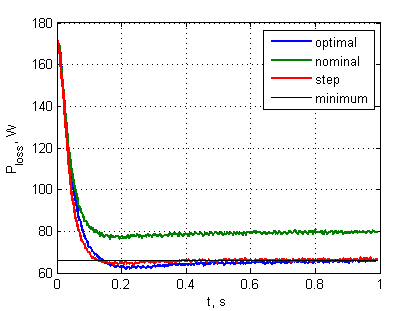}
\fi

Figure 12. Measured $P_{loss}$ for load torque step decrease from 6.8 Nm to 13.6 Nm at 60 rad/s for 4 kW motor.
\end{center}

\begin{center}
Table 8.
Measured transient energy for load step change from 6.8 Nm to 13.6 Nm at speed 100 rad/sec.

\begin{tabular}{ | l | c | c | c | c | c | }
\hline 
Method & $J_{rise}$, 100 rad/s & $J_{fall}$, 100 rad/s \\
\hline 
nominal & 171.74  & 81.39   \\
optimal & 169.93  & 68.45  \\
step & 167.57  &  68.99  \\
\hline 
\end{tabular}

\end{center}

\begin{center}
Table 9.
Measured transient energy for load step change from 6.8 Nm to 13.6 Nm at speed 60 rad/sec.

\begin{tabular}{ | l | c | c | c | c | c | }
\hline 
Method & $J_{rise}$, 60 rad/s & $J_{fall}$, 60 rad/s \\
\hline 
nominal & 171.95  & 81.48   \\
optimal & 170.03  & 68.45  \\
step & 167.57  & 68.99   \\
\hline 
\end{tabular}

\end{center}

From data obtained following conclusions can be made:

- proposed method \eqref{eq:zeta} converges to sub-optimal value of magnetizing current $i_{sd}$ (within 4 W), error is probably caused by modelling uncertainties of magnetizing curve which is average among several characteristics,

- step change method of $i_{sd}$ is practically the same as proposed sub-optimal one \eqref{eq:zeta}, because of limited bandwidth of speed and current controllers,

- speed setpoint weakly affects the form of transients,

- proposed method \eqref{eq:zeta} is clearly reduce transient energy with comparison to holding the magnetizing current to its nominal value.

\section*{Conclusion}

The initial result of the paper is a trajectory \eqref{eq:optimal_control} which is optimal transient energy minimizer for simplified problem formulation in case of ideal speed controller performance and absence of saturation in motor. Calculations show that the proposed simplification by neglecting transient contribution of $i_{rd}$ does not lead to a significant deterioration in the quality of obtained solutions.
Also the impact of limited bandwidth of real speed controller is practically almost negligible if sufficient accuracy of speed stabilization provided. For case of main induction saturation the sub-optimal current trajectory \eqref{eq:zeta} is suggested. Numerical calculations showed that the relative accuracy of this rule is practically the same as for exact solution and increased with increasing of rated motor power. Hardware implementation and experimentation showed that developed optimal strategies are operational with real motors.

\ifCLASSOPTIONcaptionsoff
  \newpage
\fi

\begin{IEEEbiography}{Alex Borisevich}
TODO:

\end{IEEEbiography}

\begin{IEEEbiography}{Gernot Schullerus}
TODO:

\end{IEEEbiography}




\end{document}